\documentclass[preprint,epsfig,12pt]{aastex63}

\newcommand{\etal}{{\em et al.\ }}

\def\ion[#1 #2]{#1\,{\sc #2}}
\def\lamb[#1]{#1\,{\AA}}
\def\lambr[#1-#2]{{{#1}--{#2}\,{\AA}}}
\def\rat[#1 #2]{#1/#2}
\def\serts89{SERTS-89}

\def\tabul{\hbox{\raise 0.75pt\hbox{$\triangleleft$}}}
\def\ergs[#1]{#1 {ergs}~{cm$^{-2}$}\,{s$^{-1}$}\,{sr$^{-1}$}}
\def\dens[#1]{10$^{#1}$\hskip 1.5pt{cm$^{-3}$}}
\def\densr[#1 #2]{10$^{#1}$\hskip 1pt{--}\hskip .5pt{10$^{#2}$}\hskip 1.5pt{cm$^{-3}$}}
\def\fl[#1 #2]{{#1}$\pm${#2}}
\def\orb[#1 #2]{{$#1^{#2}$}}
\def\ls[#1 #2]{{$^{#1}${#2}}}
\def\tm[#1 #2 #3]{{$^{#1}${#2}$_{#3}$}}

\begin{document}

\title{Hinode/EIS coronal magnetic field measurements at the onset of a C2 flare}

\shortauthors{Landi, Hutton, Brage, Li}
\shorttitle{Hinode/EIS oronal magnetic field measurements at the onset of a C2 flare}

\author{E. Landi}
\affil{Department of Climate and Space Sciences and Engineering, University of Michigan}

\author{W. Li}
\affil{Key Laboratory of Solar Activity, National Astronomical Observatories, Chinese Academy of Sciences, 20A Datun Road, Beijing 100101, People's Republic of China}

\author{T. Brage}
\affil{Department of Mathematical Physics, Lund University, Post Office Box 118, SE-22100 Lund, Sweden}

\author{R. Hutton}
\affil{Institute of Modern Physics, Fudan University, Shanghai 200433, People's Republic of China}

\begin{abstract}
In the present work we study Hinode/EIS observations of an active region taken
before, during and after a small C2.0 flare in order to monitor the evolution 
of the magnetic field 
evolution and its relation to the flare event. We find that while the flare left
the active region itself unaltered, the event included a large Magnetic Field
Enhancement (MFE), which consisted of a large increase of the magnetic field to 
strengths just short of 500~G in a rather small region where no magnetic field 
was measured before the flare. This MFE is observed during the impulsive phase 
of the flare at the footpoints of flare loops, its magnetic energy is sufficient 
to power the radiative losses of the entire flare, and has completely dissipated
after the flare. We argue that the MFE might occur at the location of the 
reconnection event triggering the flare, and note that it formed within 22 
minutes of the flare start (as given by the EIS raster return time). These 
results open the door to a new line of studies aimed at determining whether 
MFEs 1) can be flare precursor events, 2) can be used for Space Weather 
forecasts; and 3) what advance warning time they could allow; as well as 
to explore which physical processes lead to their formation and dissipation, 
whether such processes are the same in both long-duration and impulsive 
flares, and whether they can be predicted by theoretical models.
\end{abstract}


\section{Introduction}

Solar flares are one of the most important manifestations of solar activity, consisting
of the sudden release of vast amounts of energy in small portions of active regions, 
which heat the coronal plasma to ten million degrees or more and increase the plasma 
density in a matter of minutes. The high temperature and density
cause the plasma X-ray emission to increase even by orders of magnitude during the very 
first phase of the flare (the {\em impulsive phase}) and then to fade back to pre-flare
values during the {\em decay phase} as the plasma temperature decreases with time (e.g.
Fletcher \etal 2011, Benz 2016 and references therein); after which, enhanced emission 
in the EUV range outlasts the decay phase in what is called the {\em EUV late phase} 
(Woods \etal 2011, Chen \etal 2020).

The amount of energy released through radiation ranges between 10$^{27}$ to 10$^{31}$ 
erg (Ryan \etal 2012), and is capable of altering the ionization of planetary atmospheres 
and of harming astronauts working outside spacecraft with radiation poisoning. Enhanced 
ionization in the Earth's upper atmosphere in turn causes significant alterations in 
the propagation of radio waves and thus significantly affects navigation systems and 
communication, and alters the trajectories of artificial satellites by increasing 
ionospheric densities in the dayside (Pulkkinen 2007 and references therein). Also, 
flares are often (but not always) associated to Coronal Mass Ejections (CMEs, Webb \&
Howard 2012), where vast amount of plasma are released and accelerated away from the 
solar corona and can reach and interact with planetary magnetosphere, causing their 
own array of damages and disruptions to human assets on the ground (power grids) and 
in space (artificial satellites) (Pulkkinen 2007).

Because of all these negative effects, flares are one of the major components of 
Space Weather, and the prediction of their occurrence and strength is one of the 
most active fields of research in solar physics. However, CME arrival at
the Earth can be predicted either by the combination of remote sensing observations 
of their onset at the Sun and modeling of their propagation in the Heliosphere, or 
by direct in-situ detection by spacecraft at L1, such as WIND (Acuna \etal 1995)
and ACE (Stone \etal 1998). Thus, in
principle, CME mitigation strategies may dispense from a true prediction of the
occurrence of a CME event at the Sun, and focus on propagation properties of CMEs
that have been seen erupting in the solar corona. Such a strategy is not possible
with flares, as their Space Weather product only consists of radiation, so that 
any advance warning of their occurrence must rely on truly predicting capabilities.

There have been many attempts at identifying precursors of flare occurrence, both 
with using observables such as line broadening (Harra \etal 2001, 2009) or pre-flare 
brightenings (Fletcher \etal 2011), or with machine learning (Jiao \etal 2020, Wang
\etal 2020), but so far success is limited.
One of the main obstacles to flare forecasting is that the main agent that lies
at the heart of the flare phenomenon has proved to be very elusive to detection.
In fact, it is now commonly accepted that flares occur when a sudden reconfiguration 
of the coronal magnetic field in an active region structure takes place, caused by 
magnetic reconnection, so that the energy released from a flare ultimately comes 
from the magnetic field in the active region. However, measuring the magnetic
field of the solar corona has proved to be very difficult, due to the weakness
of its signatures on remote sensing observables: spectral lines. In fact,
spectropolarimetry has been capable of measuring magnetic field orientation in
the plane of the sky and line-of-sight component only at the solar limb, while
radio observations have provided measurements on disk active regions only in a 
limited number of cases. Additional magnetic field estimates from wave properties 
are difficult to use for flare predictions (Landi \etal 2020 and references therein).

As a result of the difficulty at measuring the coronal magnetic field, magnetic 
reconnection has never been unambiguously identified, a direct association of a 
flare to the coronal magnetic field in the host active region has not yet been 
made, and the amount of 
flare energy has never been compared to a measurement of energy lost by the 
coronal magnetic field. Also, no study is available to determine whether any 
change in the magnetic field itself prior to the flare event can be used as 
a precursor sign of an imminent flare.

Recently, Landi \etal (2020) developed a new diagnostic technique that allows the
direct measurement of the strength of the magnetic field from high spectral and
spatial resolution observations of a handful of bright \ion[Fe x] and \ion[Fe xi]
coronal lines
from the EUV Imaging Spectrometer (EIS, Culhane \etal 2007) on board the Hinode 
satellite (Kosugi \etal 2007). The brightness of these lines make the monitoring 
of active regions potentially hosting a flare possible, to follow the evolution 
of the coronal magnetic field before a flare.  

In this paper, we analyze a series of observations of an active region hosting 
a flare, and determine the evolution of the magnetic field before, during and 
after the flare. Section~\ref{methodology} reviews the magnetic field strength 
measurement techinque, and Section~\ref{data} describes the data availability
and the observations we utilied. Magnetic field measurements are reported in
Section~\ref{results} and discussed in Section~\ref{conclusions}.

\section{Methodology}
\label{methodology}

The diagnostic technique we use to measure the strength B of the coronal magnetic 
field has been introduced by Landi \etal (2020 -- L20), which improved on a previous 
technique first introduced by Si \etal (2020a and 2020b), and based on an idea first 
suggested a few years ago (Li \etal 2015, 2016). This technique 
capitalizes on the properties of the metastable \ion[Fe x] \orb[3s 2]\orb[3p 4]\orb[3d ] 
\tm[4 D 7/2] level which, in the presence of an external magnetic field, can mix 
with the \orb[3s 2]\orb[3p 4]\orb[3d ] \tm[4 D 5/2] level, acquiring a new decay 
channel to the ground \orb[3s 2]\orb[3p 5] \tm[2 P 3/2] level through a Magnetic-field
Induced Transition (MIT). MITs have been observed in other atomic systems when the 
external field exceeds 1~T, but they can be observable in \ion[Fe x] at much lower magnetic 
fields due to the very small energy difference between the two \tm[4 D 5/2,7/2] 
levels. As a result, the intensity of the M2 spectral line emitted by the 
\tm[4 D 7/2]-\tm[2 P 3/2] transition under normal conditions is enhanced by 
the MIT emission: the intensity of the MIT line can match and exceed the intensity 
of the M2 line already at B$\approx$200~G, well within the range of active region 
field strengths; for example, at 900~G the MIT rate is already 10 times larger 
than the M2 rate.

What makes this \ion[Fe x] property so important is that the \tm[4 D 5/2,7/2]-\tm[2 P 3/2]
transitions provide one of the brightest spectral lines in the EIS wavelength range,
easily observed at 257.26~\AA\ since the start of the Hinode mission in 2007. Thus,
the excess MIT emission can be used to measure the magnetic field strength B in 
active regions with high cadence observations. The \ion[Fe x] line observed by EIS
at 257.26~\AA\ is made by an unresolved blend between the M2 and MIT 
\tm[4 D 7/2]-\tm[2 P 3/2] transition on one side, and the \tm[4 D 5/2]-\tm[2 P 3/2] 
E1 transition on the other. L20 identified two regimes for magnetic field measurement
(see L20 for details) from the 257.26~\AA\ line:

\begin{enumerate}

\item The {\em Weak Field} regime: when B$<$200~G the MIT/M2 branching ratio 
can be determined from the intensities of the 257.26~\AA\ and another bright 
\ion[Fe x] line and a density diagnostics, and provides the most accurate B 
values. The $A_{MIT}/A_{M2}$ branching ratio

\begin{equation}
\frac{A_{MIT}}{A_{M2}} = \frac{I_{MIT}}{I_{M2}} = \frac{I_{257}}{I_{184}}R{\left[{184/M2}\right]}-R{\left[{(E1+M2)/M2}\right]}
\label{weak_field}
\end{equation}

can be calculated using the CHIANTI database and measured EIS intensities, and 
be compared with the values of the MIT/M2 branching ratio estimated as a function 
of the magnetic field by Li \etal (2021).

\item The {\em Strong Field regime}: when B$>$200~G the same lines can be used
as the weak field regime, to calculate the ratio

\begin{equation}
\frac{I_{M2+MIT}}{I_{184}} = \frac{I_{257}-I_{184}\times R{\left[{E1/184}\right]}}{I_{184}}=
\frac{I_{257}}{I_{184}}-R{\left[{E1/184}\right]}
\label{strong_field}
\end{equation}

This ratio can be calculated with the CHIANTI database as a function of the magnetic 
field and compared with observations.

\end{enumerate}

\noindent
In Equations~\ref{weak_field} and \ref{strong_field}, $I_{257}$ and $I_{184}$ are 
the observed intensities of the 257.26~\AA\ and 184.54~\AA\ \ion[Fe x] lines, and 
$184, M2, E1$ are the CHIANTI values for the contribution functions of the 
184.54~\AA, the $M2$ and $E1$ transitions. More details on the derivation of
Equations~\ref{weak_field} and \ref{strong_field} can be found in L20.

It is important to note that the M2, E1 and 184 contribution functions are density
sensitive, so an independent estimate of the electron density is needed. L20 identified
as the best density sensitive ratio to be the \ion[Fe xi] 182.17/(188.22+188.30) line 
ratio, as all other available \ion[Fe x] are either intrinsically weak or lie in 
regions of the EIS detector where instrumental sensitivity is very low.

L20 also indicated that this diagnostic technique was sensitive to magnetic fields
larger than $\approx$50~G at active region densities, and 10-20~G at quiet Sun 
densities.

The main limitation of the L20 technique is currently given by the EIS instrument 
calibration, whose accuracy L20 estimated to be $\approx 50\%$. Such an accuracy 
severely limits the applicability of the L20 technique in the strong field regime,
as the uncertainties allow it to determine only an upper limit to the magnetic
field strength. On the contrary, the weak field regime is able to provide 
measurements, albeit with a 70\% uncertainty. The reason for this is that the
intensity ratio used in the Strong Field regime (Equation~\ref{strong_field})
has a weaker dependence on the magnetic field than the ratio used the Weak Field
regime (Equation~\ref{weak_field}), so that the effect of uncertainties is much
worse in the Strong Field regime. For this reason, we will utilize the Weak Field
regime in all cases, even though, as discussed in L20, the magnetic field values
larger than 200~G are somehow underestimated.

\section{Data}
\label{data}

\subsection{The EIS observing sequence}

In order to determine the evolution of the magnetic field of an active region before,
during and after a flare, we needed to find EIS observations fulfilling {\em all} the 
following criteria:

\begin{enumerate}

\item Include a flare;

\item Observe the host active region for enough time before and after the flare occurs;

\item Include both the \ion[Fe x] 257.26~\AA\ and 184.54~\AA\ lines;

\item Include the \ion[Fe xi] 182.17~\AA\ and 188.22+188.30~\AA\ line pair, or another 
line pair from \ion[Fe ix], \ion[Fe xi] or \ion[Fe xii] allowing an accurate density 
measurement, provided it is emitted by the same plasma structures as \ion[Fe x];

\item Observe with a reasonably high cadence;

\item Image the region where reconnection occurs.

\end{enumerate}

\noindent 
The two most difficult criteria to satisfy at the same time are the first two: in fact,
the flare watch sequences routinely utilized by the EIS team to monitor active regions
for flares, which have provided a database of spectroscopically resolved flare observations,
do not include the 257.26~\AA\ line so the L20 technique can not be applied. Many other
flares have been observed serendipitously in active regions while executing large rasters
lasting hours, and usually only once or twice, so that no monitoring is possible.

L20 listed around 90 observing sequences that include the two \ion[Fe x] needed for
magnetic field diagnostics, and we have inspected all the individual observations 
utilizing each of these rasters in the EIS archive from the beginning of the EIS 
mission to the time of this writing. Unfortunately, we were able to identify only 
one data set that satisfied all the above criteria. These are the data we have 
analyzed in this work.

The observations were taken early in the EIS mission, on 24-Aug-2007, over a small active
region at the beginning of the solar cycle 24 minimum. The observing sequence is called 
{\em PRY\_footpoints\_v2} and includes 22 spectral windows scattered across the EIS 
wavelength range, which allow the measurement of a large number of spectral lines both 
in the transition region and in the corona; also, the \ion[Fe xxiii] 263.89~\AA\ and 
\ion[Fe xxiv] doublet at 192.03~\AA\ and 255.11~\AA, as well as a few other high 
temperature lines are included.

The field of view of the observation is 100"$\times$240" and each raster position was
observed for 25s, for a total raster duration of 22m17s. It was pointed at AR10969
and observed it for 7 consecutive times from 6:29:17~UT to 8:42:59~UT with no 
interruption.

\section{Results}
\label{results}

\subsection{The flare}

During the fourth raster (started at 7:36:08~UT) the Geostationary Operational 
Environment Satellites (GOES) X-ray monitor detected
a small impulsive C2.0 flare initiating at 7:49~UT, peaking at 7:54~UT and ending 
at 7:58~UT. At the time, the EIS spectrometer was rastering across a portion of the 
host active region, as the intensity of the flare \ion[Fe xxiii] and \ion[Fe xxiv]
peaked. As the flare progressed, the EIS slit moved eastward across a small loop-like 
flare structure, which then faded in the background. No sign of any flare activity
was present in the following raster passing over the same region hosting this structure.

\subsection{The active region hosting the flare}

\begin{figure}
\includegraphics[height=20.0cm,width=15.0cm,angle=90]{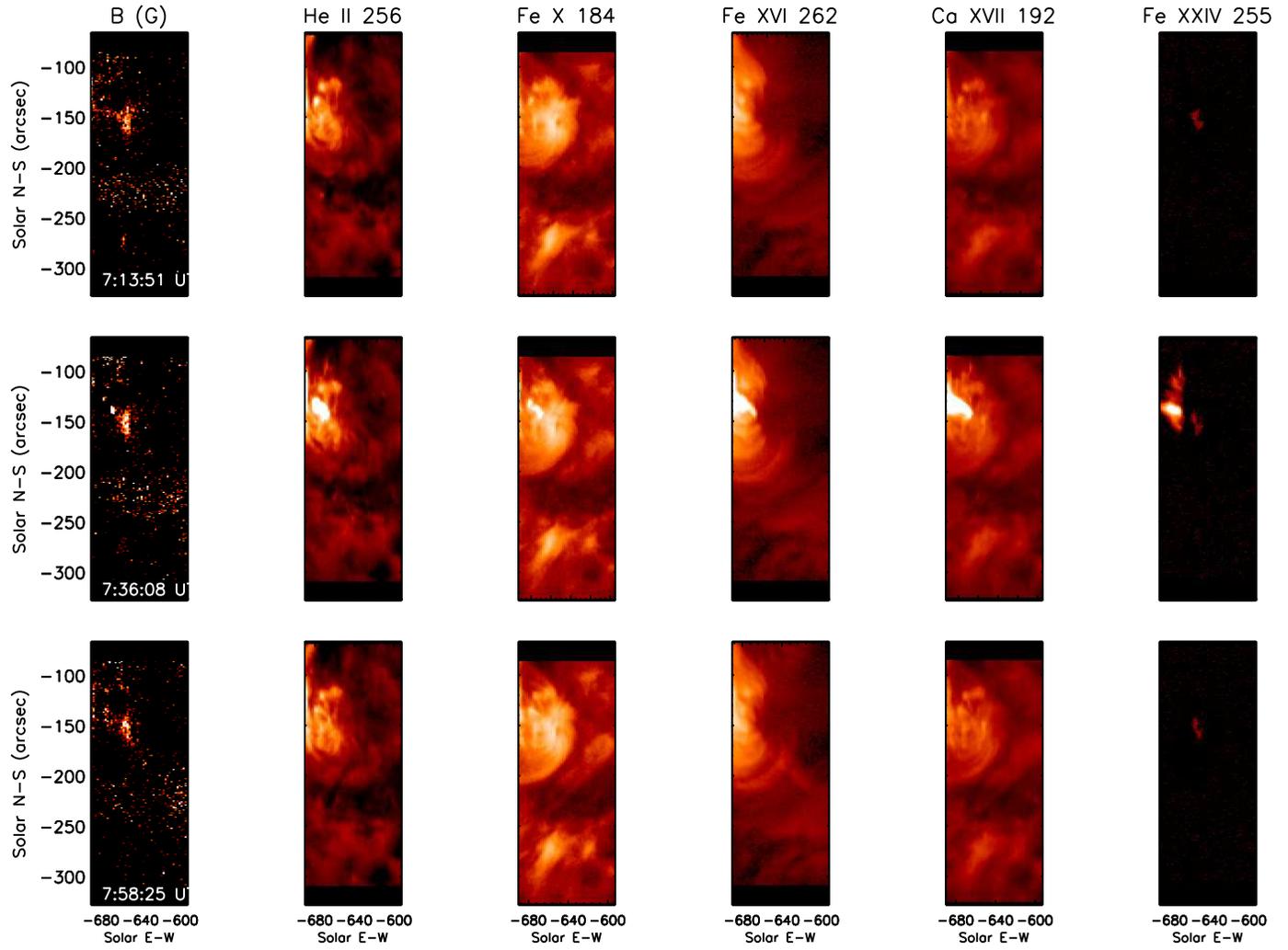}
\caption{Intensity and magnetic field strength maps for three consecutive EIS
rasters of AR10969, taken before, during and after the C2 flare. The flare can
be seen in the 07:36~UT raster (middle row).
\label{maps_time}}
\end{figure}

Figure~\ref{maps_time} shows intensity maps as measured in five different spectral 
lines for each of three rasters across the active region, centered on the raster 
hosting the flare. The lines range in formation temperature from the chromosphere 
(\ion[He ii] 256.32~\AA), to the corona (\ion[Fe x] 
184.4~\AA\ and \ion[Fe xvi] 262.98~\AA), and the flare (\ion[Ca xvii] 192.85~\AA\ 
and \ion[Fe xxiv] 255.11~\AA). The magnetic field strength map is also shown in the 
first column, showing that only in the region at the base of a coronal loop structure 
was the L20 magnetic field diagnostic technique able to detect magnetic field of 
strong enough to provide a significant measurement. 

\begin{figure}
\includegraphics[height=20.0cm,width=20.0cm,angle=90]{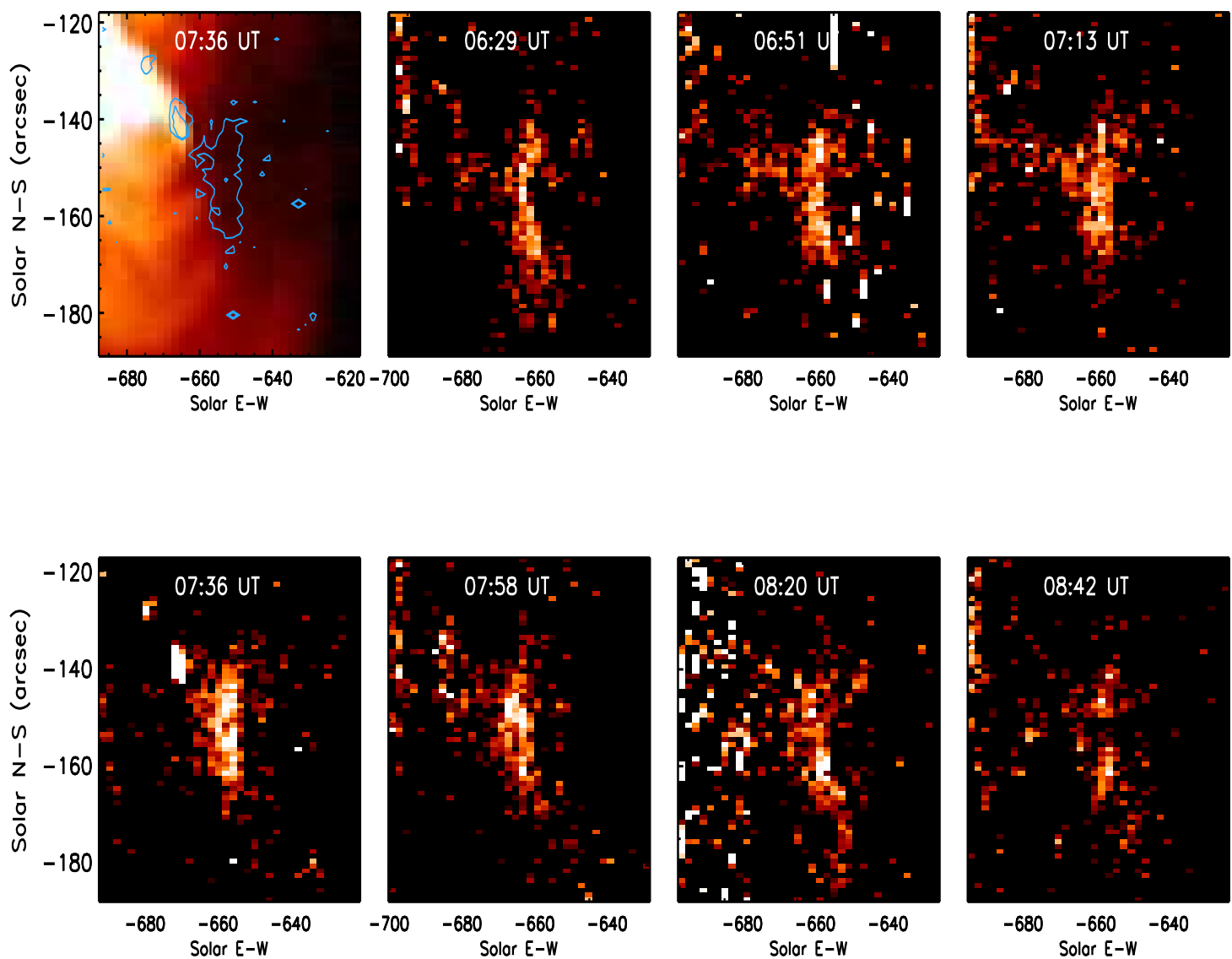}
\caption{Maps of the magnetic field overlaying the sunspot in each of the
first 7 observations. For reference, the intensity of the \ion[Fe xvi] 262.98~\AA\
line at 07:36~UT is shown in the top left panel, with the countours of the magnetic
field measured at the same time overlaid on it. The MFE is present in the 07:36~UT
(bottom-left) panel only and is located at the footpoint of the \ion[Fe xvi] flare
emission at 07:36~UT (top left panel).
\label{b_sunspot}}
\end{figure}

\begin{figure}
\includegraphics[height=18.0cm,width=12.0cm,angle=90]{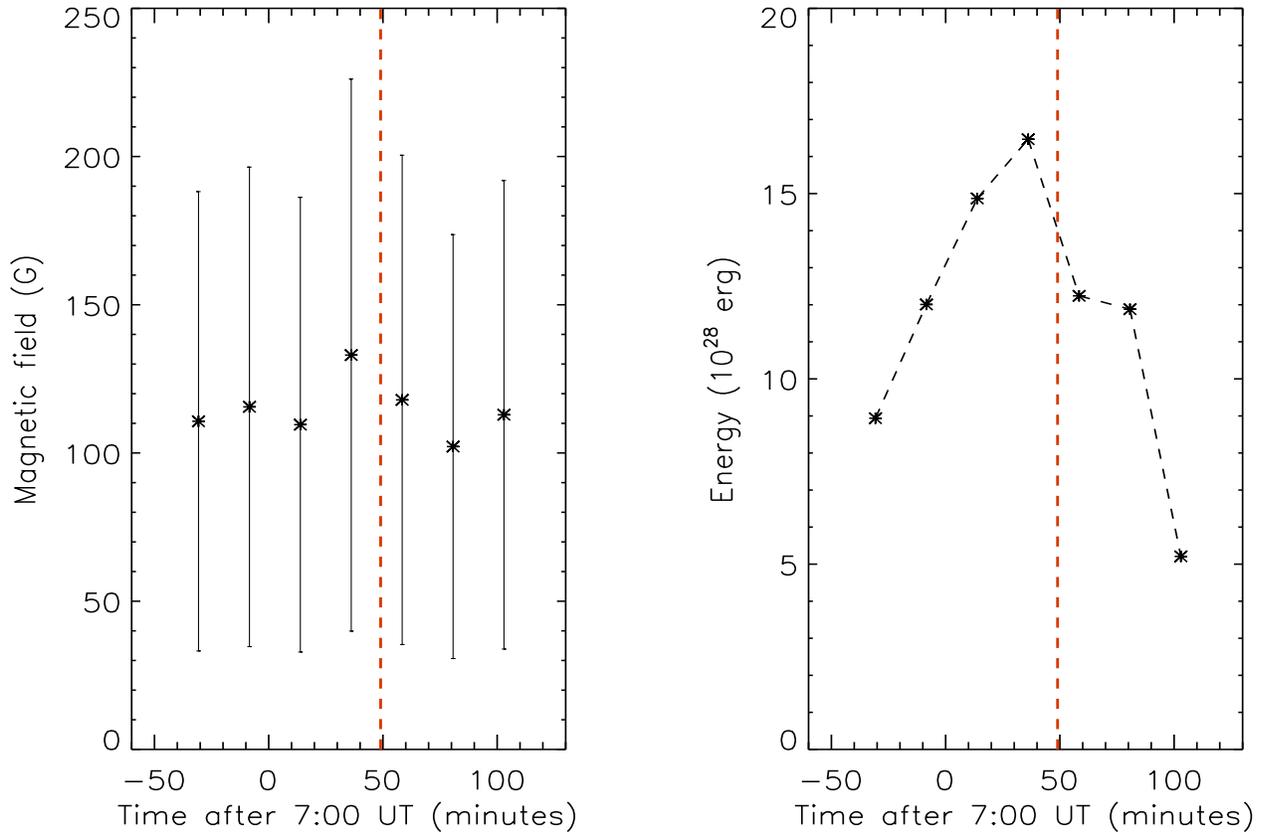}
\caption{Average magnetic field overlying the sunspot, as shown in Figure~\ref{b_sunspot},
and magnetic energy associated to it.
\label{b_average}}
\end{figure}

A closer view of the evolution of the magnetic field in the active region is given 
in Figures~\ref{b_sunspot} and \ref{b_average}; in the latter, the flare start time 
measured by GOES is indicated by the vertical red dashed line. In Figure~\ref{b_sunspot}, 
the morphology of the magnetic field is displayed, showing that the shape and distribution 
of the magnetic field significantly changed, with the size of the region with magnetic 
field stronger than $\approx$50~G slightly increasing before the flare (occurred in
the 07:36~UT raster) and decreasing afterwards. 

The average magnetic field strength, shown in the left panel of Figure~\ref{b_average}, 
does not change significantly and is about 100~G, with the large uncertainty erasing 
the possibility of detecting any evolution. On the contrary, the total magnetic energy 
(right panel) shows some evolution, largely due to the change in size of the region 
filled with detectable magnetic field. This energy has been calculated (in erg) as

\begin{equation}
E_{tot} = 10^7\frac{B^2}{2\mu_0}V
\label{energy}
\end{equation}

\noindent
where $B$ is the magnetic field strength (in T), and $\mu_0$ is the permeability of
free space. The volume $V$ of the region has been determined by multiplying the area 
(in $m^2$) of the image occupied by magnetic field larger than 50~G in each raster 
(which is the quantity that causes the energy variation in Figure~\ref{b_average}, 
right) with an arbitrary average line-of-sight (LOS) depth of the magnetic field 
structure imaged by the \ion[Fe x] emission of 1" (corresponding to 734~km, with
a solar radius of 949" as seen from the Earth). It is important to stress that these
energy values are only gross estimates, whose uncertainties stem from two main
sources: 1) the uncertainty in the value of $B$ (around 70\%, as suggested in L20);
and 2) the complete lack of information about the LOS depth of the magnetic field 
region imaged by \ion[Fe x].

\begin{figure}
\includegraphics[height=20.0cm,width=20.0cm,angle=90]{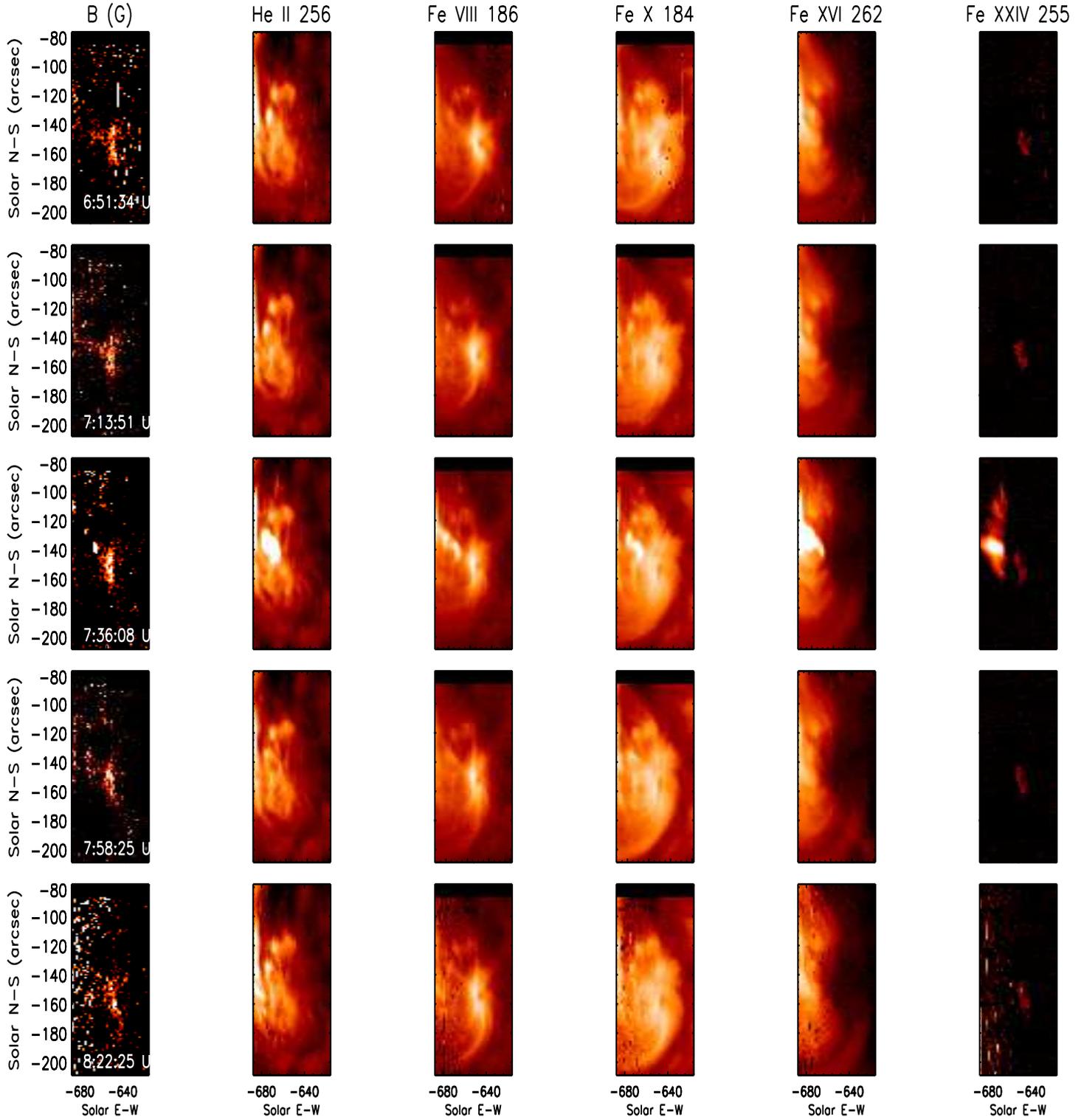}
\caption{Close-ups of the magnetic field strength and intensity maps of the flaring 
active region.  The flare occurred at 07:49~UT, and is observed in the 07:36~UT 
observation (middle row).
\label{active_region_time}}
\end{figure}

The evolution of the whole active region before, during and after the flare is shown
in more detail in Figure~\ref{active_region_time}. Except for the appearance of the 
\ion[Fe xxiv] line in the 07:36~UT image, invisible in all other rasters, along with 
a brightening of variable intensity and shape in all other lines at the same time, 
the overall structure of the active region is largely untouched by the event. Also, 
the flare has no significant effect on the magnetic field with the only exception 
of a large magnetic field enhancement over a very small area in the 07:36~UT raster 
at around ${\left({-670",-140"}\right)}$. It is fair to say that the flare left the 
host active region relatively unscathed.

\subsection{The flare Magnetic Field Enhancement}

\begin{figure}
\includegraphics[height=19.0cm,width=16.0cm,angle=90]{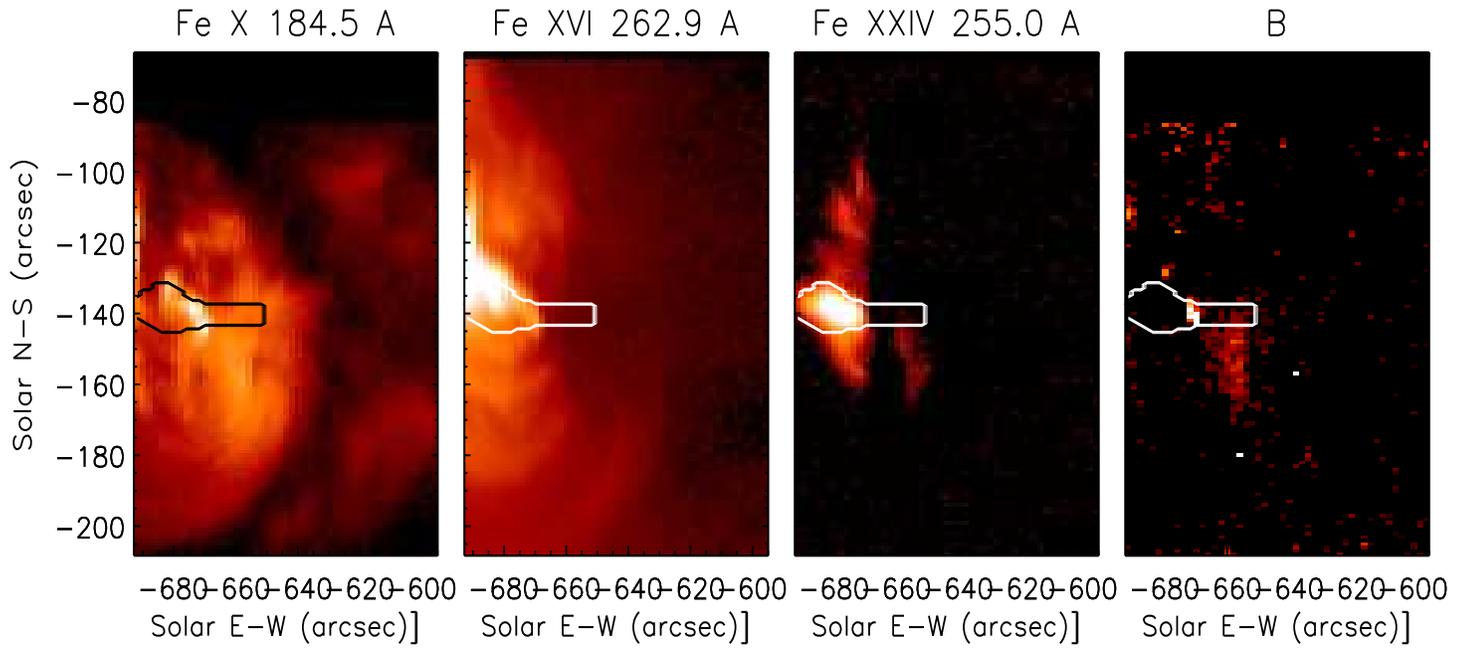}
\caption{Magnetic field and intensity maps of the flaring active region from the
07:36~UT observation, with the pixels selected for diagnostics overlaid on
them. The MFE is visible at two slit raster positions
at (-670",-140").
\label{flare_map}}
\end{figure}

The magnetic field enhancement (MFE) and its relationship to the flare emission 
are shown in the top left 07:36~UT panel in Figure~\ref{b_sunspot}, where it is 
overlaid (blue line) to the \ion[Fe xvi] flare intensity taken at the same time. 
While the MFE seems to be unrelated to the active region magnetic field, it is 
located right at the footpoint of the \ion[Fe xvi] flare intensity enhancement; 
Figure~\ref{flare_map} confirms that this point lies at the footpoint of the 
flare emission of all lines, especially the \ion[Fe xxiv] emission.

In order to measure the magnetic field strength and the flare properties, we have
selected the pixels along the slit that include the flare emission, marked using
the \ion[Fe xxiv] 255.11~\AA\ line, and the MFE, and extended the selection to 
pixels westward of the flare, at locations observed before the flare erupted, 
for comparison purposes. The selected pixels are shown in Figure~\ref{flare_map}. 
For each slit position, the emission has been averaged along the slit and used
used to measure line intensities, plasma densities and the magnetic field strength.

\begin{figure}
\includegraphics[height=17.0cm,width=22.0cm,angle=90]{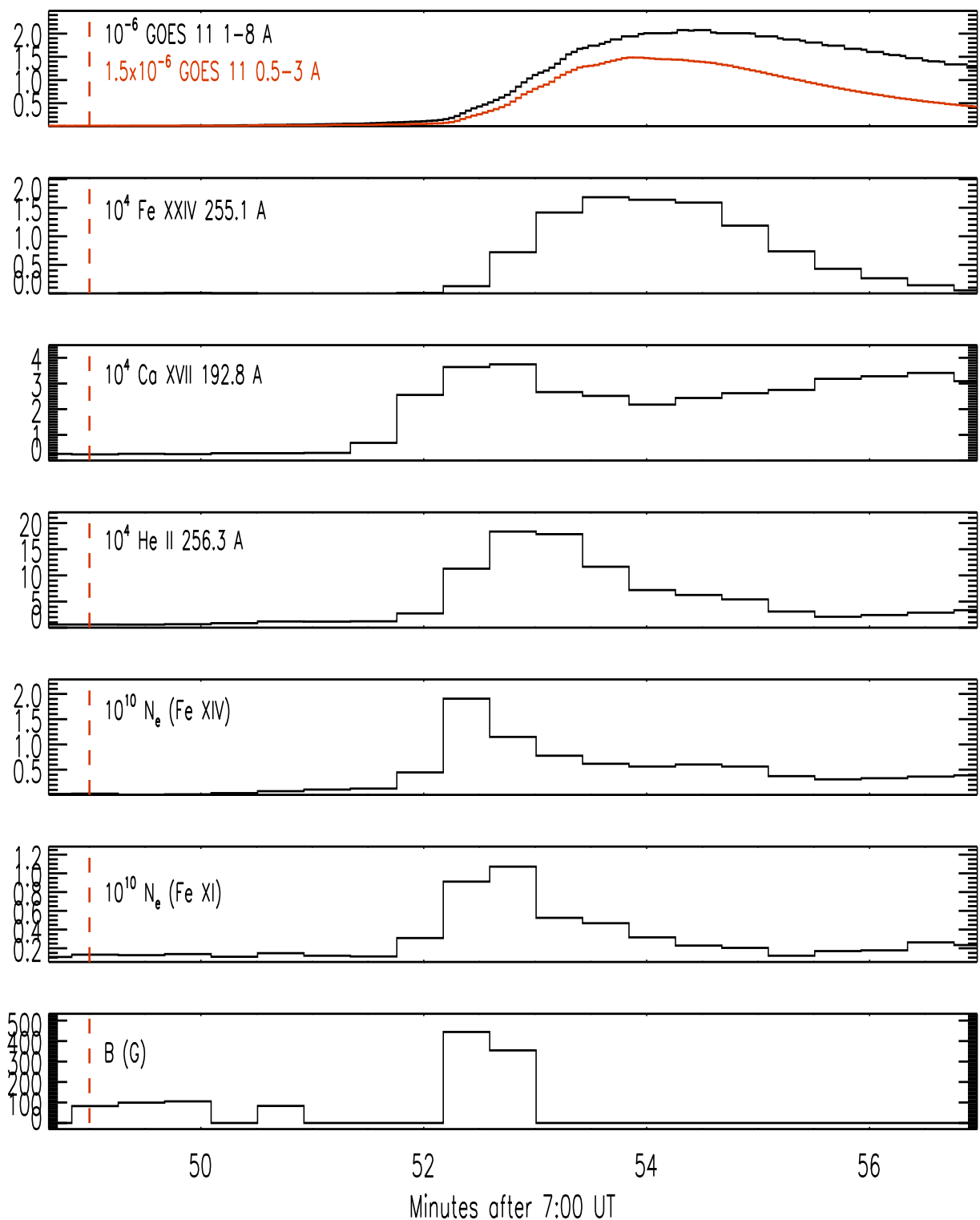}
\caption{Light curves observed during the flare, from top to bottom: GOES-11 X-ray 
flux, the \ion[Fe xxiv] 255.11~\AA\ and
\ion[Ca xvii] 192.85~\AA\ flare lines, the \ion[He ii] 256.32~\AA, the electron density
from \ion[Fe xiv] and \ion[Fe xi] intensity ratios, and magnetic field strength B
measured with the L20 diagnostic technique. The start of the GOES flare is marked by
the red vertical dashed line.
\label{flare_timing}}
\end{figure}

Figure~\ref{flare_timing} shows the light curves of GOES-11, the \ion[Fe xxiv] 
255.11~\AA\ and \ion[Ca xvii] 192.85~\AA\ flare lines, the \ion[He ii] 256.32~\AA\ line, 
and the electron density measured using the \ion[Fe xiv] 264.24/274.20 and \ion[Fe xi] 
182.17/(188.22+188.30) line intensity ratios, along with the magnetic field measurement.
The start of the flare as reported by GOES is indicated by the vertical red dashed line
at 07:49~UT, on the left. What is extremely important to remember is that the rastering 
nature of the EIS observing sequence means that the EIS light curves are built using 
data taken as the slit was rastering, so that different times correspond to different 
locations within the active region, which might have been evolving differently from each 
other. Thus, a direct comparison between the GOES X-ray light curve and the EIS light 
curves is potentially misleading, as GOES was observing the Sun as a star and therefore 
includes the emission from the entire flare. Also, time variations in any of the 
EIS-related quantities might be equally due to either the flare time evolution or 
its spatial variability.

With this caveat in mind, Figure~\ref{flare_timing} allows us to draw several conclusions. 

\begin{itemize}

\item First, EIS did not observe the location of the reconnection event triggering the flare, 
or at least reached it after reconnection occurred, as the flare effects were imaged by 
the instrument a few minutes later than the GOES start, indicating that the flare
took place eastward of the location observed by EIS at 07:49~UT. 

\item Second, EIS started
to observe significant flare emission and density enhancement around 2 minutes after the 
flare started, corresponding to 6 steps (12~arcsec) in the E-W direction: since the X-ray 
emission was still rapidly increasing when EIS reached the flare location, the energy 
release was still underway. A further confirmation of this comes from the \ion[Ca xvii] 
light curve rising and peaking earlier than \ion[Fe xxiv], as expected in a rapidly
heating plasma, unless multiple locations of energy injections were present.

\item Third, the plasma density increases significantly from pre-flare values, with 
\ion[Fe xi] and \ion[Fe xiv] providing different estimates, indicating that plasma 
at different temperatures may belong to different structures below the EIS spatial 
resolution affected by the flare.

\end{itemize}

\begin{figure}
\includegraphics[height=17.0cm,width=22.0cm,angle=90]{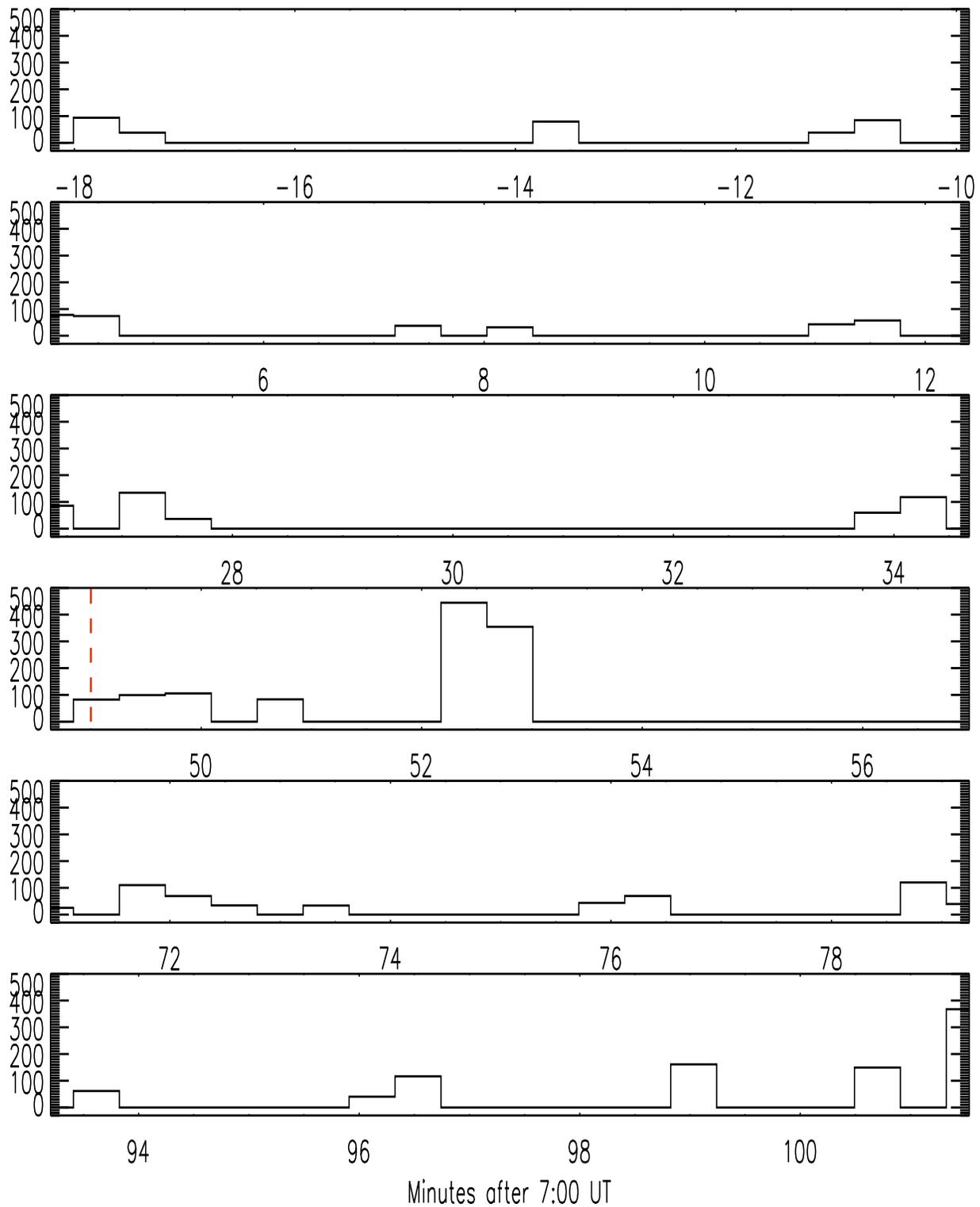}
\caption{Magnetic field strength measured from the same pixels used to study 
the flare, along 6 consecutive EIS rasters, from 6:29~UT to 8:41~UT.
Each time stamp on the X-axis of each row corresponds to a different pointing; however,
the locations shown are the same for each consecutive raster.
\label{b_time}}
\end{figure}

However, the most stricking feature of Figure~\ref{flare_timing} is the presence of
a strong MFE right at the locations where the line emission is rising and the electron 
density has spiked. This MFE is all the more striking as it occurs in a region where 
the magnetic field is below the detection threshold of the L20 technique when observed 
before the flare, and its spatial extent is very limited, being around 8~arcsec along 
the slit and just two raster positions (corresponding to 4~arcsec) along the E-W 
direction. This MFE is observed {\em after} the flare has already started (as shown 
by the GOES light curve, and even the \ion[Ca xvii] intensity enhancement) and 
{\em before} the impulsive phase ends and energy is still being released. Also, this 
MFE is present {\em only} in the raster containing the flare: Figure~\ref{b_time} 
shows the magnetic field strength measured in 6 consecutive rasters at the same 
pixels selected for the flare, and clarly shows that no magnetic structure is 
present at the MFE location in the rasters before and after the flare. 

Finally, the energy associated to this MFE is, according to Equation~\ref{energy}, 
around $8\times 10^{28}$~erg, again assuming arbitrarily that LOS length of the 
region where the magnetic field is enhanced is 1". This energy is in good agreement 
with the typical total energy lost to radiation by a C2 flare, which lies in the 
$10^{28}-10^{30}$~erg range (Ryan \etal 2012). It is important to note that this 
energy has been obtained from measurements using the Weak Field regime, as discussed in 
Section~\ref{methodology}, and thus represents a underestimate of the total 
energy. Still, despite the very large uncertainty in the actual value of the 
magnetic energy of the MFE, this measurement clearly shows that the MFE detected 
by the L20 technique is capable of providing enough energy to power the entire 
flare.

All these properties of the MFE point towards the following conclusions:

\begin{enumerate}

\item The MFE most likely takes place at the location of the magnetic reconnection event;

\item The MFE is strictly related to the flare event, having developed 
during the 22 minutes that EIS took to come back to the same location;

\item The MFE energy is sufficient to power the entire flare;

\item The enhanced magnetic field is dissipated during the flare to levels below
the L20 technique threshold.

\end{enumerate}

\section{Discussion}
\label{conclusions}

The results of the present analysis indicate that a short impulsive C2 flare, with no 
visible effects on the structure of the host active region, is strictly associated to 
a relatively strong MFE (350-500~G) that formed right before the flare event, and was 
entirely dissipated during the flare event. This MFE is confined to a relatively small 
region (16$\times$2~arcsec$^2$) where no specific plasma structure was visible before 
or after the flare, and leaves no trace after the flare ends. The magnetic energy 
associated to the MFE is compatible with typical total radiative losses from a C2 
flare, and thus it is capable to power the flare itself. 

These results open a lot more questions than they answer, and pose a formidable challenge 
to flare prediction efforts. Hopefully new observations from the EIS spectrometer will be 
devoted to flare hunts and will lead to a database of flares of all sizes that allow to 
improve on the present results and extend them to all flare classes. 

The first conclusion that we can draw about flare predictability is that the quick formation
of an MFE in an active region might be a precursor sign of a flare. While this provides a 
potential powerful tool for flare prediction, the short time that the MFE took to form (less 
than 22 minutes, corresponding to the return time of the EIS raster) poses a formidable 
challenge to flare prediction efforts, as it indicates that only a few minutes are available 
between the detection of the flare precursor formation and the actual flare.

Still, a C2 flare is a minor flare, and the properties of larger flares, with significant
impact on Space Weather (M1.0 and higher) can be different, such as taking more time to
form an MFE with enough energy to generate a more powerful flare. Furthermore, impulsive 
and long-duration flares might have different MFE formation processes and timescales.

The fact that we could only find one suitable flare dataset in the entire EIS range means
that our results are just one point in the flare parameter space. They open the door to
a host of future studies, which hopefully will answer the many questions which the present 
results can only ask:

\begin{enumerate}

\item Are MFEs a typical flare feature?

\begin{enumerate}
\item What processes are responsible for MFE formation, and where within the active region
do they form?
\item How much time does an MFE need to form, and is this formation time correlated to
the flare strength?
\item Can MFE formation itself be predicted?
\item Do all impulsive and long-duration flare produce MFEs, and are the formation 
processes and locations the same?
\end{enumerate}

\item Can MFEs be used as flare precursors for Space Weather forecasting? 

\begin{enumerate}
\item If MFEs are indeed flare precursors, which flare prediction success rate would
they allow us to reach?
\item What advance warning time can MFEs give us for a flare event?
\end{enumerate}

\end{enumerate}

\noindent
Also, many issues about the L20 technique itself need to be addressed in order to make the
estimate of the magnetic field strength and energy more accurate. These issues mainly concern
the calibration of the EIS instrument, which is the main source of uncertainty in the 
determination of an absolute value of the MFE magnetic field strength and energy.

\acknowledgements

E. Landi was supported by NSF grants AGS 1408789, 1460170, and NASA grants NNX16AH01G, 
NNX17AD37G and 80NSSC18K0645.

\centerline{\bf References}
\begin{enumerate}

\item Acuna, M.H., Ogilvie, K.W., Baker, D.N., \etal 1995, Sp. Sci. Rev. 71, 5
\item  Benz, A. 2017, Liv. Rev. Sol. Phys., 14, 2
\item  Chen, J., Liu, R., Liu, K., \etal 2020, ApJ, 890, 158
\item Culhane, J.L., Harra, L.K., James, A.M., \etal 2007, Sol. Phys., 243, 19
\item  Fletcher, L., Dennis, B.R., Hudson, H.S. \etal 2011, Sp. Sci. Rev, 159, 19
\item  Harra, L.K., Matthews, S.A., \& Culhane, J.L. 2001, ApJ, 549, L245
\item  Harra, L.K., Williams, D.R., Wallace, A.J., \etal 2009, ApJ, 691, L99
\item Jiao, Z., Sun, H., Wang, X., \etal 2020, Sp. Weather, 18, e2020SW002440
\item Kosugi, T., Matsuzaki, K., Sakao, T., \etal 2007, Sol. Phys., 243, 3
\item Landi, E., Hutton, R., Brage, T., \& Li, W. 2020, ApJ, 904, 87
\item Li, W., Grumer, J., Yang, Y., \etal 2015, ApJ, 807, 69
\item Li, W., Yang, Y., Tu, B., \etal 2016, ApJ, 826, 219
\item Li, W., Li, M.,Wang, K., Brage, T., Hutton, R., Landi, E. 2021, ApJ {\em submitted}
\item  Pulkkinen, T. 2007, Liv. Rev. Sol. Phys, 4, 1
\item  Ryan, D.F., Milligan, R.O., Gallagher, P.T., \etal 2012, ApJS, 202, 11
\item Si, R., Brage, T., Li, W., \etal 2020a, ApJL, 898, L34
\item Si, R., Li, W., Brage, T., \& Hutton, R. 2020b, J. Phys. B, 53, 095002
\item Stone, E.C., Frandsen, A.M., Mewaldt, R.A., \etal 1998, Sp. Sci. Rev., 86, 1
\item Wang, X., Cheng, Y., Toth, G., \etal 2020, ApJ, 895, 3
\item  Webb, D.F., \& Howard, T.A. 2012, Liv. Rev. Sol. Phys., 9, 3
\item  Woods, T.N., Hock, R., Eparvier, F., \etal 2011, ApJ, 739, 59

\end{enumerate}

\end{document}